\documentclass[twocolumn,floats,aps,amsmath]{revtex4}
\usepackage{rotating}
\usepackage{graphicx}
\usepackage{dcolumn}
\usepackage{bm}
\usepackage{amssymb} 
\usepackage{amsmath} 
\usepackage[utf8]{inputenc} 

\begin{document}

\preprint{APS/123-QED}

\title{Detection of high-field superconducting phase in CeCoIn$_5$
with magnetic susceptibility}
\author{S.D. Johnson}
\affiliation{Physics Department, University of California Davis, Davis, CA
95616}
\author{R.J. Zieve}
\affiliation{Physics Department, University of California Davis, Davis, CA
95616}
\author{J.C. Cooley}
\affiliation{
Los Alamos National Laboratory, Los Alamos, NM 87545}

\date{\today}

\begin{abstract}

We measure the ac susceptibility of single-crystal CeCoIn$_5$ in dc field
parallel to the {\em c} axis and find further evidence for a high-field
phase transition within the superconducting phase in this orientation.
We apply up to 2.3 kbar uniaxial pressure along the $c$ axis and discuss
the pressure dependence of the high-field phase.  We also report the
behavior of $H_{c2}$ under uniaxial pressure for field along $c$.

\end{abstract}


\maketitle

The heavy-fermion superconductor CeCoIn$_5$ and its isostructural cousins
Ce(Ir,Rh)In$_5$ and Pu(Co,Rh)Ga$_5$ exhibit several intriguing features.
Collectively these compounds have ground states that are superconducting,
antiferromagnetic, or both. The materials have significant two-dimensional
character, suggested by their layered crystal structure and confirmed
through electronic structure calculations.  Furthermore, within each
family $T_c$ varies linearly with the ratio of lattice constants $c/a$,
illustrating how closely geometry and superconductivity can be linked
\cite{pagliuso}.  With a $c$-axis magnetic field CeCoIn$_5$ also has a
quantum critical point near the upper critical field, which can be shifted
into the superconducting phase with hydrostatic pressure \cite{ronning}.

Yet another unusual feature appears in the high-field, low-temperature
portion of the superconducting regime.  This has been studied mainly for
magnetic fields applied in the basal plane, where $H_{c2}$ is about 11.8
Tesla and a signature within the superconducting phase is seen near 10
Tesla with magnetization \cite{tayamaPRB}, specific heat \cite{miclea},
neutron diffraction \cite{kenzelmann}, and NMR \cite{young} measurements.
The additional signature moves to higher fields as temperature increases,
intersecting the $H_{c2}(T)$ curve near 300 mK.  These data are of
particular interest since CeCoIn$_5$ is a good candidate to support the
Fulde-Ferrell-Larkin-Ovchinnikov (FFLO) superconducting phase described
almost 50 years ago \cite{fulde, larkin}. This state is marked by a
non-zero quasiparticle center-of-mass momentum and a spatially periodic
superconducting order parameter.  CeCoIn$_5$ is in the extremely
clean limit and has strong Pauli limiting of its critical field. In
addition, the high-field phase (HFP) is accompanied by a change of the
normal-superconducting transition from second-order at low fields to
first-order at high fields, in agreement with FFLO theory \cite{tayamaPRB,
bianchifo,radovan}.
While the exact nature of the phase and its identification with FFLO
remain in question \cite{kenzelmann, young}, the experimental evidence
for the HFP with in-plane field is clear.

\begin{figure}[bh]
\begin{center}
\scalebox{.42}{\includegraphics{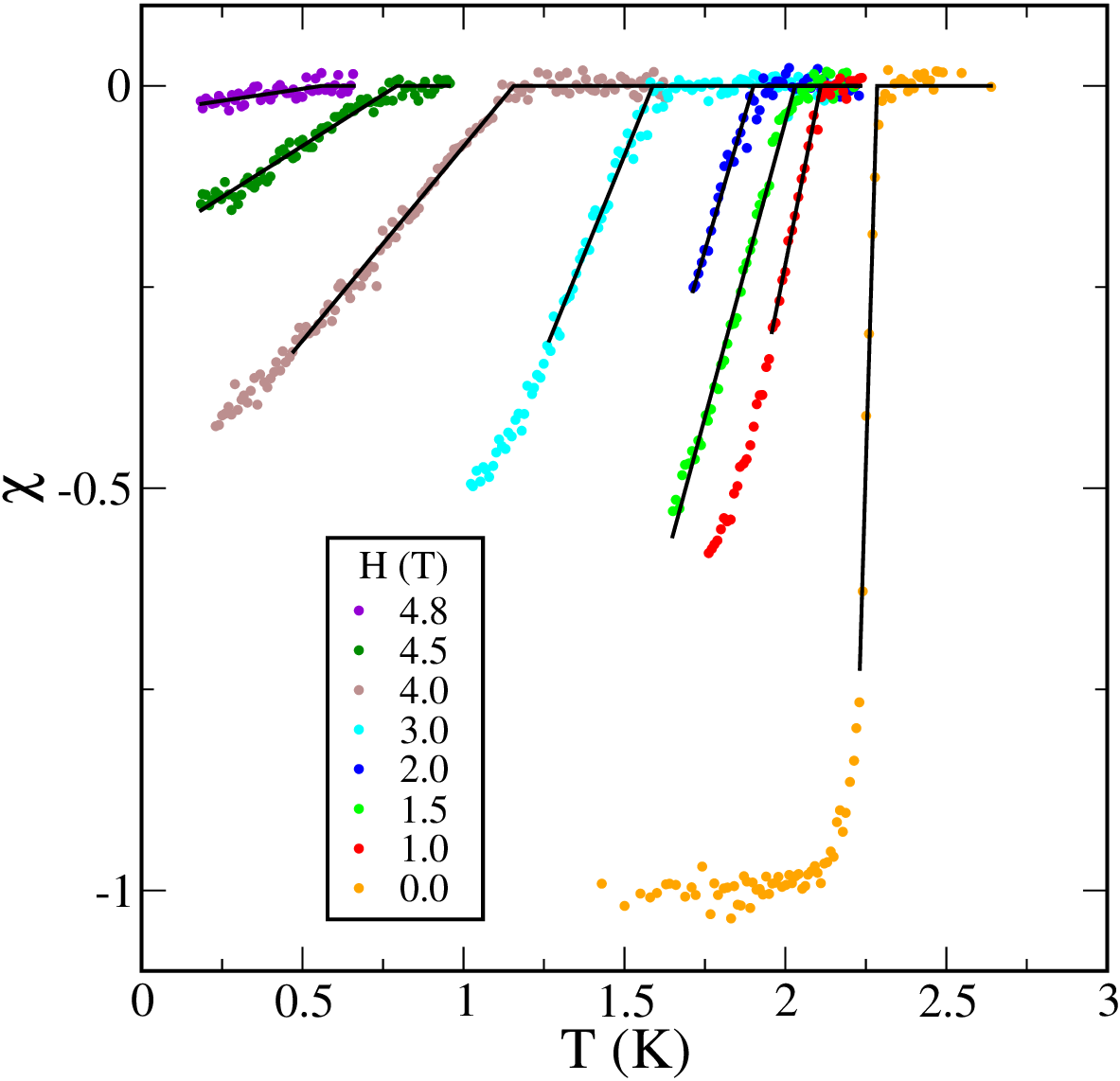}} 
\caption[Data at the onset of $T_c$ for different dc fields and at fixed
pressure with overlaid fits.]{Onset of
superconductivity for different dc fields and at fixed pressure of 1.82 kbar,
with overlaid fits.  Legend indicates the applied field in Tesla
for each curve.}
\label{onset}
\end{center}
\end{figure}

Much less effort has been devoted to CeCoIn$_5$ with magnetic field
along the $c$ axis. A few measurements have indicated a high-field
phase for this orientation too, as seen both by signatures within
the superconducting phase \cite{bianchihc, kumagai, gratens} and by a
change of the normal-superconducting transition from second-order
to first-order similar to that observed for in-plane field.  However,
other experiments have found no signatures \cite{bianchifo, miclea},
leaving even the experimental situation unclear.  In this report we
provide further evidence for the HFP phase in CeCoIn$_5$ for $H || c$,
along with the uniaxial pressure dependence of the HFP boundary up to
2.32 kbar.

We measured three samples oriented with uniaxial pressure, dc field,
and ac field parallel to the \emph{c} axis.  The pressure is applied
with a bellows setup activated with helium gas from room temperature.
The samples had mass of 3.67 mg, 0.67 mg and 0.92 mg with area 2.78
$\times 10^{-6}$ m$^2$, 5.16$\times 10^{-7}$ m$^2$, and 6.97 $\times
10^{-7}$ m$^2$ respectively.  The experimental details, including sample
preparation and orientation, are identical to those we reported previously
\cite{johnson}, with the exception of the additional dc field used in
this study.

For four runs in the pressure cell at nominally zero pressure and zero
field, we found an average $T_c=2.29$ K.  In each case the transition
temperature measured in the pressure cell was within 24 mK of the
value outside the pressure cell.  In some cases the temperature shift can
indicate the initial pressure applied during the experimental setup and
cooling process, typically about 0.3 kbar \cite{dix,johnson}.
However, $T_c$ in CeCoIn$_5$ is not very sensitive to $c$-axis pressure,
shifting by less than 17 mK/kbar \cite{johnson, oeschler}, 
so it gives little information on the pressure offset.
Hence the values for applied pressure used in this paper do not
include any offset for an initial pressure.

Figure \ref{onset} presents susceptibility curves near the onset of
superconductivity for several applied fields, all at a pressure of 1.82 kbar.
As indicated by the black lines in the figure, we determine $T_c$ by
assuming the susceptibility is constant in the normal state and changes
linearly just below $T_c$.  As usual for superconductors, the onset $T_c$
shifts to lower temperature as the applied field increases.  In addition,
the transition decreases in magnitude and broadens in temperature.
The magnitude change is due to flux lines threading the sample, which
lead to incomplete diamagnetism.  The broadening is governed by the
temperature-dependence of the vortex pinning.  As temperature falls and
the vortices are more strongly pinned, the behavior becomes increasingly
diamagnetic.  We now turn to temperature sweeps below 400 mK and at fields
above 4.5 Tesla.  In this regime the susceptibility has not saturated but
retains a temperature-dependence which decreases with increasing field.

We confirmed that changing the pressure causes no systematic
shift in either $\chi_s$ or $\chi_n$. Nor does field shift $\chi_n$. Together
these indicate that only $\chi_s$ is affected and only by tuning applied field. 
This corresponds to a decrease in superconducting transition magnitude with
increasing field. 

\begin{figure}[tb]
\begin{center}
\scalebox{.45}{\includegraphics{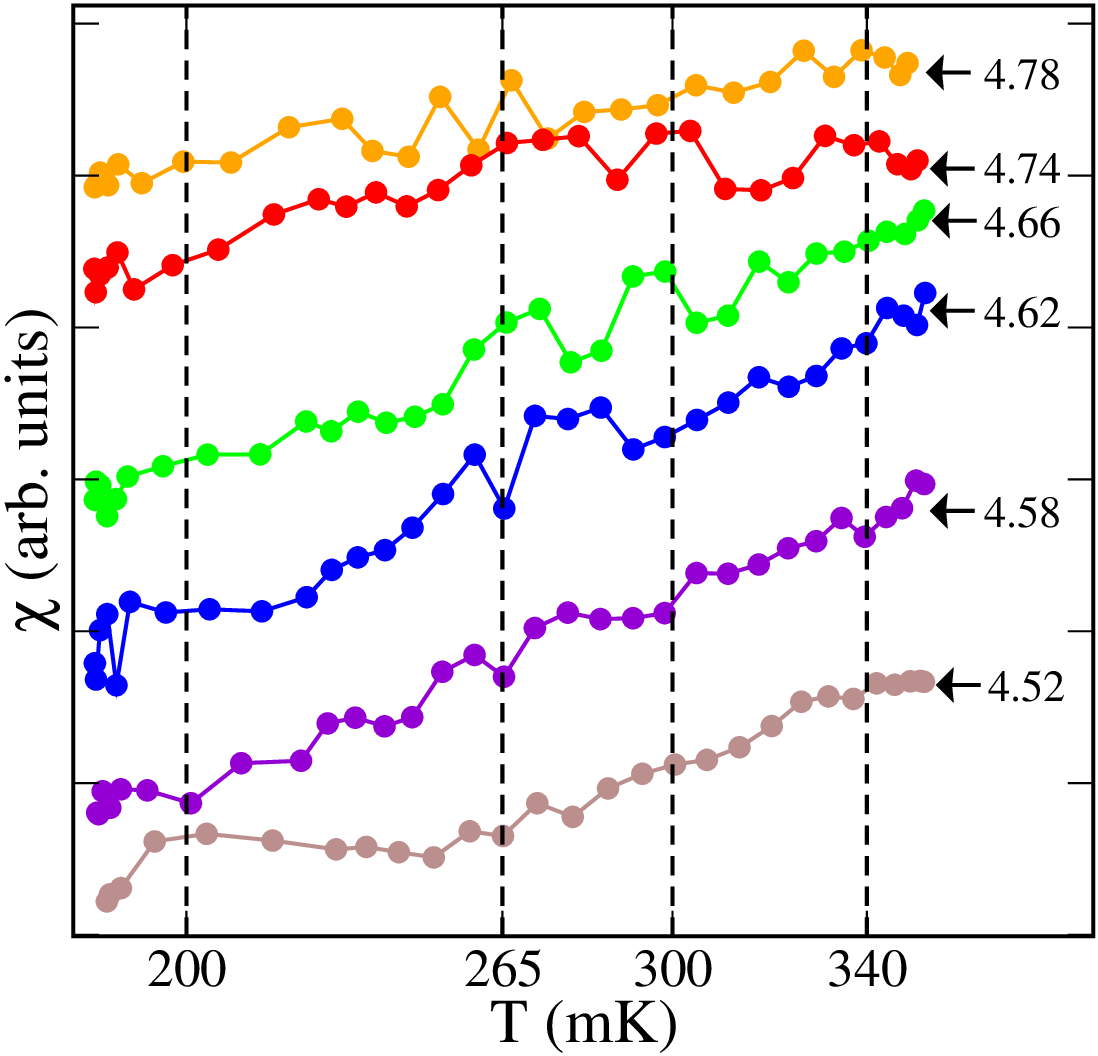}} 
\caption{Susceptibility deep in the superconducting phase at
1.42 kbar.  The applied field for each curve is indicated to the right.  The
vertical dashed lines indicate the temperatures that we use to set up
the field vs temperature curves as shown in
Figure \ref{kink}.}
\label{scfits}
\end{center}
\end{figure}

\begin{figure}[tb]
\begin{center}
\scalebox{.45}{\includegraphics{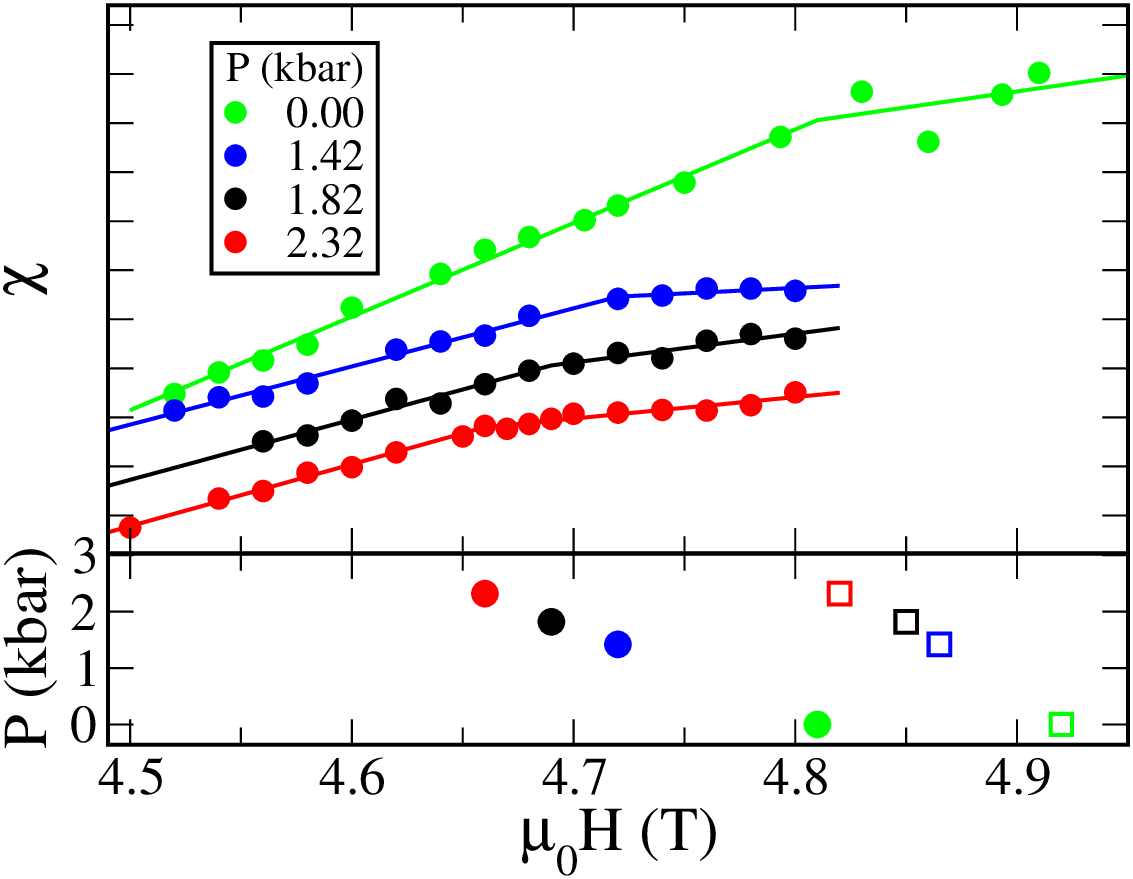}} 
\caption{Top: Susceptibility at 300 mK with fits overlaid.  See
text for details. All data except zero pressure are from the same sample and
are shifted vertically for clarity.  Bottom: Plot of the location of
the kinks $H_k$ obtained from the fits (filled circles), with $H_{c2}$
at 300 mK also shown (open squares).}
\label{kink}
\end{center}
\end{figure}

Susceptibility at much lower temperatures appears in Figure \ref{scfits}.
The data shown were measured at 1.42 kbar uniaxial pressure.  When we
increase the dc field the superconducting susceptibility $\chi_s$
decreases in magnitude.  We track this shift
as a function of field by picking out $\chi$ at a fixed temperature
from each of these dat sets.  Figure \ref{kink} shows some of the resulting
$\chi$ vs $H$ curves.
We repeat this procedure for each of the four temperatures indicated
by the dashed lines of Figure \ref{scfits}.  In each case we smooth the
data by averaging $\chi$ in a 40 mK range centered at the desired temperature.

The upper frame of Figure \ref{kink} shows data at 300 mK for several
different pressures.  Each data set appears to have linear regimes,
but also a kink where the slope changes abruptly.  At each pressure we
fit a function of exactly this form: two linear portions, with a kink at
($H_k$,$\chi_k$) where the slope changes.  We use four free parameters:
$H_k$, $\chi_k$, and the slope on each side.  We vary $H_k$ and $\chi_k$
manually, in steps of 0.01 Tesla and 0.00075, respectively.  For each
kink location, we solve for the slopes which best fit the data above and
below $H_k$.  We then select the kink location which minimizes the least
squares error. The lines overlaid in the center plot are the result of
this minimization.  The lower plot indicates the location of $H_k$ and
also the critical field $H_{c2}$ at 300 mK for each pressure.  The two
decrease in concert with increasing pressure, showing that most of
the kink's shift with pressure comes from the reduction of $H_{c2}$.

\begin{figure}[tb]
\begin{center}
\scalebox{.42}{\includegraphics{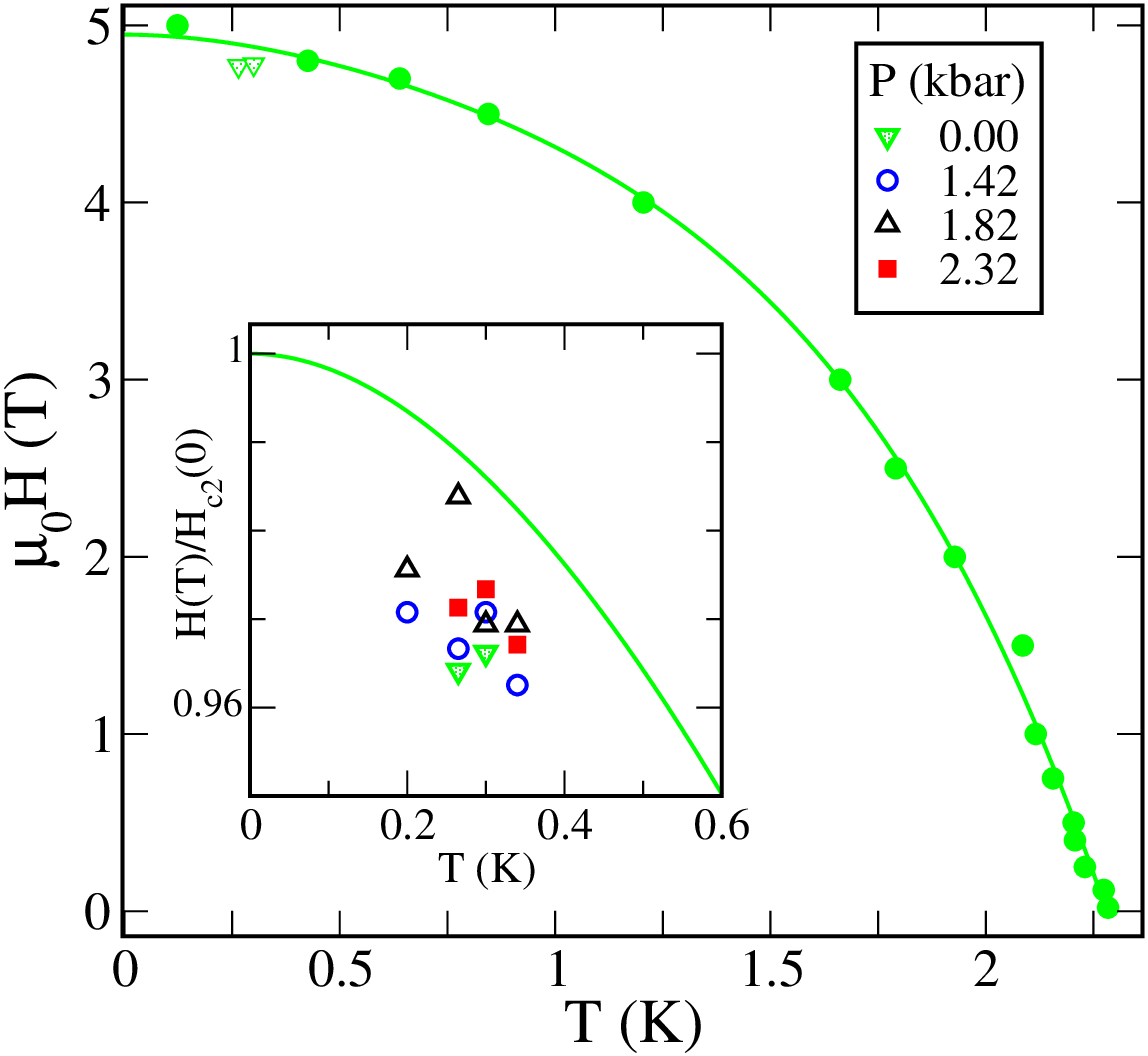}} 
\caption{Main graph presents measurements of $H_{c2}(T)$ at
zero pressure (solid circles) and a fit to these points.  Shaded triangles
indicated the kink locations $H_k(T)$.  
Inset shows an expanded view in the region of $H_k$ near the top of
the dome.  The curve $H_{c2}(T)$ is normalized to 1 at $T=0$, and $H_k$
is shown for several applied pressures.
}
\label{phase}
\end{center}
\end{figure}

The main portion of Figure \ref{phase} shows $H_{c2}(T)$ at ambient
pressure, with $H_k$ also plotted.  As seen in the lower part of Figure
\ref{kink}, $c$-axis pressure reduces $H_{c2}$ at low temperatures.
At 500 mK, the change in $H_{c2}(T=0)$ is 0.04 T/kbar.  Hydrostatic pressure
has a somewhat larger effect, between 0.06 and 0.08 T/kbar \cite{miclea,
tayama}.  Interestingly, although increasing hydrostatic
pressure also pushes the high-temperature, low-field portion of the $H_{c2}(T)$
dome up to higher $T_c$, this shift is almost absent with uniaxial
pressure.  We recently reported that the zero-field $T_c$ has little
dependence on $c$-axis pressure, rising only about 20 mK to a maximum
near 2 kbar, so it is not surprising that $c$-axis pressure also
has little effect on the phase boundary at very low fields.

The inset of Figure \ref{phase} expands the high-field, low-temperature region. 
The solid curve is $H_{c2}(P=0)$, normalized to 1 at $T=0$.  The
normalized $H_{c2}$ curves at other pressures are nearly identical and are
omitted from the graph.  The symbols show the normalized kink field,
$H_{k}(T,P)/H_{c2}(0,P)$. The kink agrees well
with the proposed HFP boundary, which appears between 0.955 and 0.96 in
other measurements \cite{bianchi, matsuda, kenzelmann,young}.
Several transitions of the vortex lattice have also been observed
\cite{bianchi}, but the highest of these is near 0.85, much lower
than the kink.  Under uniaxial pressure both $H_{c2}$ and $H_k$
decrease, by roughly the same factor.
$H_k$ is also approximately independent of temperature and forms
a linear boundary well below $H_{c2}$.

To complement the temperature sweeps described above we also performed field
sweeps at fixed temperature and pressure.  One might expect a kink as in Figure
\ref{kink} to appear directly in this measurement.  However, in fact we
find no evidence of a kink.  At all fields below $H_{c2}$, the slope of
$\chi$ vs $H$ in the field sweeps is comparable to its high-field value
in the temperature-sweep data.  The low-field slope in the temperature
sweeps is notably larger.  The sensitivity to the measurement history
could explain why some of the previous experiments have found no sign
of a transition to the HFP for $c$-axis field.

One possible source of the kink is influence from the AFM QCP,
which at ambient pressure occurs at a field very close to $H_{c2}$
\cite{ronning}.  However, under hydrostatic pressure the QCP field
decreases about five times as fast as $H_{c2}$, so that the QCP moves
deep inside the superconducting phase.  While its behavior has not
been tracked under uniaxial pressure, there is no reason to believe
that $H_{c2}$ and $H_{QCP}$ would remain close together.  Since we find
that $H_k$ changes with uniaxial pressure only about 1.5 times as fast
as $H_{c2}$, the kink is probably not an indication of the QCP.

Another interpretation is in terms of thermally activated flux
flow \cite{beek}.  Our temperature sweeps are nearly field-cooled
measurements. Although in practice we often do not exceed $T_c$, we find
no apparent difference in the low-temperature susceptibility between
these temperature sweeps and those when the sample does begin in the
normal state.  The larger low-temperature slope for temperature sweeps
indicates that the vortex pinning is stronger after field-cooling.
A plausible explanation is that at higher temperatures the vortices
fluctuate enough to find the most favorable pin sites, which they cannot
do when the field changes at low temperature.  In the HFP, there is no
such annealing effect, and the pinning strength always has the smaller value.

A change in pinning strength could arise from antiferromagnetic order in
the region above $H_k$, which could interact with the superconductivity.
More generally, any change in the order parameter can reasonably be
expected to affect the vortex pinning.  Several reports have been made
along with theoretical descriptions suggesting this phase is the FFLO
phase (see reference \cite{matsuda} for an overview).  In the FFLO state
periodic planar nodes appear perpendicular to the flux lines, leading to
a segmentation of the vortices into pieces of length $\Lambda = 2\pi/q$,
where $q$ is the quasiparticle wave vector \cite{matsuda} which factors
into the order parameter $\Delta \propto \sin{\bf{q} \cdot \bf{r}}$.
The pieces are relatively free of each other and hence better able
than conventional vortices to position themselves at pin centers.
Hence pinning vorces on the flux lines should increase in the FFLO phase.
If the nodal planes reflect the crystal structure, then uniaxial
pressure parallel to the flux lines would compress the segmentation
length $\Lambda$ and change the periodicity of the order parameter.
Already strongly Pauli-limited, CeCoIn$_5$ under uniaxial pressure
becomes even more so than with hydrostatic pressure, possibly due to
increased hybridization between the Ce-In layers.  The way that $H_k$
tracks $H_{c2}$ suggests a possible connection with the electron spin
population.  Theoretical models suggest redistribution of spin states in
the vortex cores upon entering the FFLO state \cite{mizushima}. Namely,
at the intersection of the nodal plane and the vortex core, excess spin
states are emptied.

Ultrasound measurements of the HFP boundary \cite{watanabe} for in-plane
field can also be interpreted in terms of vortex pinning.  The vortices
affect the ultrasound velocity more strongly in the HFP, which is
consistent with increased pinning; one explanation is segmentation of
the vortices into short, somewhat independent pieces that can better take
advantage of low-density pin sites.  Such segmentation could arise from
the planar order parameter nodes of an FFLO state.  By contrast, our work
suggests {\em lower} pinning in the HFP.  This may be evidence that the
HFP phases for the two field orientations are not the same.  However,
we note that the two measurements are not necessarily contradictory.
The ultrasound signal comes from the most strongly pinned vortices,
while our measurements probe the vortices that are most free to move.
In principle the HFP could support a wider range of pinning strengths
that produces both effects.

In conclusion, we have measured ac susceptibility response of CeCoIn$_5$
in an applied field, as well as under uniaxial pressure. We find evidence
for a phase boundary in agreement with previous indications of the
transition to the HFP at zero pressure.  We report a shift in this
boundary with uniaxial pressure which roughly tracks $H_{c2}$, decreasing
in field at a rate about 50\% faster than the decrease of $H_{c2}$.
For $H_{c2}$ itself, pressure depresses the critical field at low
temperatures but has little effect near $T_c$.  This differs from 
hydrostatic measurements but is consistent with the minimal
effect of uniaxial pressure on $T_c$ itself.  The location and pressure
dependence of the $H_k$ boundary provide further evidence for
the HFP phase for $H || c$ in CeCoIn$_5$. An unanswered question is why
the $H_k$ boundary shows up with temperature sweeps and not field sweeps.

\section{Acknowledgement}
This work was funded by the NSF though grant
DMR-0454869.

\end{document}